\documentclass[sigconf, nonacm]{acmart}

\usepackage{arxiv}

\usepackage{seqsplit}
\usepackage[most]{tcolorbox}
\usepackage{xcolor}
\usepackage{booktabs}
\usepackage{multirow}
\usepackage{graphicx}
\usepackage{caption}
\usepackage{tabularx}
\usepackage{enumitem}
\usepackage{cleveref}
\usepackage{tikz}

\renewcommand\vldbdoi{XX.XX/XXX.XX}
\renewcommand\vldbpages{XXX-XXX}
\renewcommand\vldbavailabilityurl{https://github.com/Jantory/llm-trained-matcher}

\newcommand{\header}[1]{\vspace{1mm}\noindent\textbf{#1}.}
\newcommand{\headerl}[1]{\vspace{1mm}\noindent\textit{#1}.}
\newcommand{\headerul}[1]{\vspace{1mm}\noindent\underline{\textit{#1}}.}

\newcommand{\phem}[1]{%
  \par\smallskip
  \noindent\textbf{#1.}\nobreak\hspace{0.5em}%
}

\begin{document}
\title{Beyond Scale and Generation: Understanding Language Model-based Entity Matching}

\settopmatter{authorsperrow=3}

\author{Zeyu Zhang}
\orcid{0009-0000-8829-0652}
\affiliation{%
  \institution{University of Amsterdam \& AUMC}
  \city{Amsterdam}
  \country{NL}
}
\email{z.zhang2@uva.nl}

\author{Xue Li}
\orcid{0000-0001-5451-1867}
\affiliation{%
  \institution{CWI}
  \city{Amsterdam}
  \country{NL}
}
\email{effy.li@cwi.nl}

\author{Iacer Calixto}
\orcid{}
\affiliation{%
  \institution{University of Amsterdam \& AUMC}
  \city{Amsterdam}
  \country{NL}
}
\email{i.coimbra@amsterdamumc.nl}

\author{Paul Groth}
\orcid{0000-0003-0183-6910}
\affiliation{%
  \institution{University of Amsterdam}
  \city{Amsterdam}
  \country{NL}
}
\email{p.t.groth@uva.nl}

\author{Sebastian Schelter}
\orcid{0000-0003-4722-5840}
\affiliation{%
  \institution{BIFOLD \& TU Berlin}
  \city{Berlin}
  \country{DE}
}
\email{schelter@tu-berlin.de}

\begin{abstract}
Entity matching identifies records that refer to the same real-world entity. Language models can be adapted to this task through bi-encoder, cross-encoder, and generative matcher architectures. However, prior studies often conflate matcher architecture with differences in model backbone, model variant (originating from varying pretraining objectives), and model size, making it difficult to isolate the sources of observed performance gains. To address this issue, we conduct a controlled factorial study spanning three matcher architectures, based on three model variants and three model sizes from the Qwen3 model family, and nine datasets, for a total of 1,215 fine-tuning runs. We complement our study with an additional evaluation of cross-dataset transferability and computational cost. 

Our experimental results show that the choice of model variant is critical for bi-encoders: embedding-oriented variants provide stronger initialization and more favorable representation geometry that is predictive of downstream matching performance. Cross-encoders retain a consistent advantage over bi-encoders because they jointly encode record pairs rather than representing each record independently, although increasing model size partially narrows this gap. Generative matchers do not universally outperform cross-encoders. Instead, their advantages are concentrated in settings involving distribution shift, such as subtle unseen differences in the record schemas and cross-dataset transfer. We further find that larger models tend to rely more heavily on shortcut learning and therefore do not necessarily yield better results. These findings clarify the factors underlying performance differences across matcher architectures and motivate future research and benchmark designs that more carefully disentangle architectural choices from other model-level factors while explicitly evaluating distribution shift and cross-dataset transferability. We release our experimental results, code, training scripts, and evaluation data at \url{https://github.com/Jantory/llm-trained-matcher}.

\end{abstract}

\maketitle


\section{Introduction}
\label{sec:introduction}

\begin{figure*}[t]
\centering
\includegraphics[width=1.0 \textwidth]{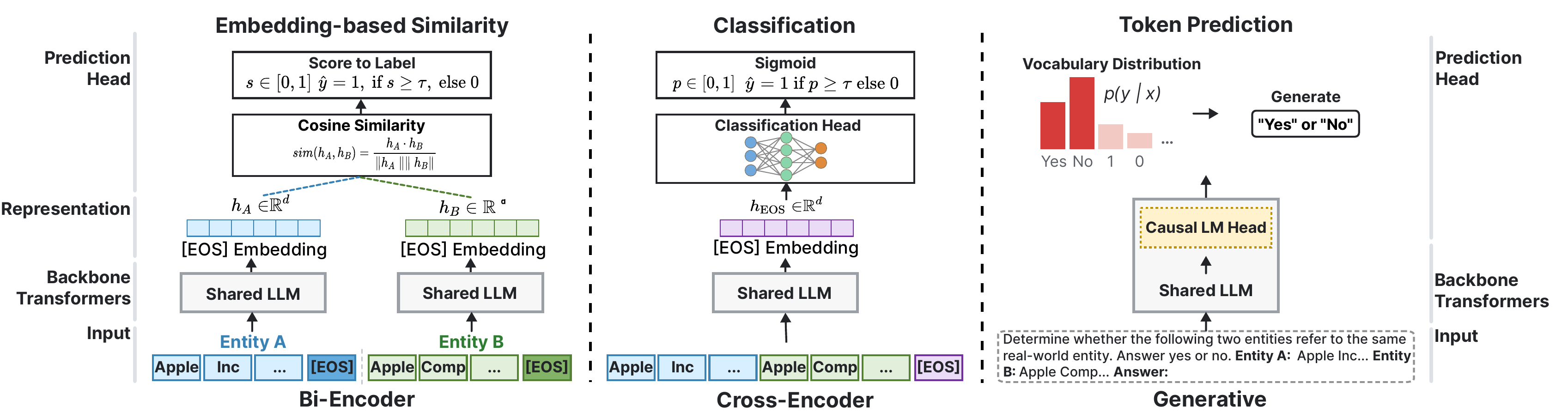}
\caption{Three LLM-based architectures for entity matching. Bi-encoders encode the two records independently and compare on representations. Cross-encoders jointly encode the record pair and predict with a classification head. Generative matchers also jointly encode the pair but retain the language-modeling head to generate a token-based decision, such as \texttt{yes} or \texttt{no}.}
\label{fig:paradigms}
\end{figure*}

Entity matching (EM) identifies data records that refer to the same real-world entity, making it a central component of data integration~\cite{brunner2020entity, dong2018data, tu2023unicorn} and knowledge base construction~\cite{zhong2023comprehensive,zhao2018architecture}. Pairwise EM is commonly formulated as a binary classification task~\cite{tu2023unicorn, mudgal2018deep}, yet the underlying decision often requires complex reasoning. For instance, a matcher must recognize semantically equivalent values expressed in different formats and align evidence across heterogeneous schemas~\cite{wang2011entity}. These requirements have made pretrained language models, and more recently large language models (LLMs), effective foundations for modern EM systems~\cite{li2020deep,zeakis2025depth,li2023effective,peeters2025entity,wang2025match,zhang2025deep}.

\header{Language model-based entity matching} However, the adoption of language models for entity matching has not produced a single standard solution. Such models can be adapted to EM through several matcher architectures that differ in how records are represented, compared, and mapped to a matching decision. Language models can be used to instantiate three main matcher architectures, as demonstrated in Figure~\ref{fig:paradigms}. A bi-encoder independently encodes two records into embeddings that serve as their representations and compares them in a shared embedding space. Because each representation can be computed without reference to any other record, record embeddings can be cached and reused. This property has made bi-encoders widely used for retrieval and blocking~\cite{wu2020scalable, zeakis2025depth}. Recent work, however, shows that with appropriate fine-tuning, they can also serve as efficient pairwise matchers~\cite{karapiperis2026impact}. A cross-encoder instead jointly encodes a record pair and directly produces a match score~\cite{li2020deep, tu2023unicorn}. Joint encoding allows the underlying Transformer layers to capture fine-grained interactions between the two records. In recent years, generative matchers based on LLMs have received increasing attention. Similar to cross-encoders, they jointly encode the record pair, but retain the language-modeling head and produce the final matching decision through conditional generation~\cite{peeters2025entity,wang2025match}.

\header{Overlooked design factors} The progression of prior work may appear to suggest an evolution from bi-encoders to cross-encoders and ultimately to generative matchers~\cite{zhang2025deep, karapiperis2026impact, peeters2025entity}. However, this apparent consensus often overlooks several design factors that are coupled with the choice of matcher architecture and therefore require systematic investigation. 
The first source of coupling concerns the underlying model backbone. Bi-encoders and cross-encoders have traditionally been built on models pretrained with masked language modeling and bidirectional attention~\cite{liu2019roberta, devlin2019bert}, whereas generative matchers rely on models pretrained with causal language modeling, in which each token attends only to preceding tokens~\cite{brown2020language}. Consequently, comparisons across matcher architectures frequently also compare different transformer architectures and attention mechanisms. A second source of coupling concerns model variant selection. Different model variants correspond to different training objectives (e.g., instruction following) before task-specific fine-tuning~\cite{yang2025qwen3}, which we refer to as pretraining objectives throughout this paper. Bi-encoders are commonly initialized from embedding-oriented models~\cite{zeakis2025depth}, cross-encoders from base models~\cite{li2020deep, tu2023unicorn}, and generative matchers from instruction-tuned models~\cite{peeters2023using, steiner2025fine}. A third source of coupling is model size. Comparisons between existing baselines often involve substantially different parameter scales. This issue is particularly pronounced in LLM-based EM, where a large generative matcher~\cite{zhang2025deep} is frequently compared against a much smaller cross-encoder such as Ditto~\cite{li2020deep}.

\header{Open research questions} These coupled design choices make it difficult to attribute observed performance differences fairly to matcher architecture itself. They can also make practical recommendations unstable: a conclusion may change when the underlying model family, variant, or  size changes~\cite{zeakis2025depth, karapiperis2026impact, peeters2025entity, zhang2025deep}. Several research gaps therefore remain. First, can task-specific fine-tuning overcome differences in initialization through a model variant, and does its effect vary across matcher architectures? Second, how large is the benefit of joint pair encoding, and can a model with stronger pretrained representations compensate for the absence of direct interaction? Third, when cross-encoder and generative matchers jointly encode the same entity pair using the same underlying backbone model, does the generative formulation still provide an advantage, and if so, under what conditions? Fourth, how do the matching architectures differ when evaluation moves from conventional within-dataset settings to cross-dataset transfer? Finally, how do training and inference costs affect architecture selection in practical deployments?

\header{Overview and main findings} To address these questions, we conduct a controlled study that disentangles matcher architecture, model variant, and model size (detailed in ~\S\ref{sec:setup}). This design is enabled by the Qwen3 model family~\cite{yang2025qwen3}, which provides models at multiple parameter sizes and offers three model variants at each scale: base, embedding-oriented, and instruction-tuned. Our core evaluation follows a full factorial design over three matcher architectures, three model variants corresponding to different pretraining objectives, three model sizes, nine datasets, and five repetitions with different random seeds. In total, the main experiment comprises \(3 \times 3 \times 3 \times 9 \times 5 = 1,215\) fine-tuning runs. We further conduct cross-dataset evaluations and systematically measure training cost, memory consumption, inference throughput, latency, and the benefits of representation reuse.

Our results provide several insights into the behavior of language model-based matchers. First, model variant selection is decisive for architectures that depend directly on representation quality, as detailed in~\S\ref{sec:three-layer:representation}. Embedding-oriented variants provide substantially stronger initialization for bi-encoder fine-tuning, and the geometry of the pretrained representation space is predictive of downstream matching performance. In contrast, the choice of model variant is not a primary determinant of cross-encoder or generative performance. Even an embedding-oriented checkpoint can be fine-tuned into an effective generative matcher.

Second, joint pair encoding provides a persistent advantage over independent encoding, as detailed in~\S\ref{sec:three-layer:interaction}. Increasing model size substantially improves bi-encoder performance, but does not eliminate the gap with cross-encoders. This result indicates that stronger independently computed representations can partly compensate for the absence of interaction, but cannot fully replace direct token-level comparison between records.

Third, generative matchers do not exhibit a universal advantage over cross-encoders, as detailed in~\S\ref{sec:three-layer:generalization}. The two architectures perform similarly on most within-dataset evaluations when using the same underlying model. Further analysis in~\S\ref{sec:three-layer:generalization} and~\S\ref{sec:cross} shows that the generative advantage is concentrated in settings involving distribution shift, including unseen subtle schema differences and cross-dataset transfer. In cross-dataset evaluation, bi-encoder behavior is strongly tied to the fine-tuning dataset and transfers poorly to heterogeneous targets. Jointly encoded matchers transfer substantially better, with generative matchers achieving the highest overall retention of target-specific performance. These results further suggest that the principal advantage of the generative formulation lies in its robustness under distribution shift rather than in conventional within-distribution pairwise matching.

Fourth, larger models do not consistently produce better matching performance, as detailed in~\S\ref{sec:three-layer:generalization}. Under shifts of the class-conditional feature distribution between the training and evaluation data, scaling can amplify reliance on shortcut learning, thereby reducing precision and F1 score. Representative and high-coverage fine-tuning data can therefore be as important as or more important than increasing model size.

Finally, the preferred matcher design depends on both the data distribution and the deployment regime, as detailed in~\S\ref{sec:cost}. Cross-encoders provide a strong effectiveness–throughput balance for supervised pairwise matching when their labeled training data adequately represents deployment conditions. Generative matchers are more attractive when robustness to distribution shift is a primary concern, although they incur higher memory costs. Bi-encoders become competitive when efficieny is a prime concern, as record representations can be cached and reused during blocking or retrieval.

\header{Contributions} In summary, our contributions are as follows:

\begin{itemize}[leftmargin=*]
\item We argue that comparisons across matcher architectures should disentangle the effect of model backbone, model variant representing different pretraining objectives, and model size, rather than attributing differences produced by these coupled design choices solely to matcher architecture (\S~\ref{sec:introduction}).

\item We conduct a controlled full-factorial evaluation across three matcher architectures, three model variants, three model sizes, nine datasets, comprising 1,215 fine-tuning runs (\S~\ref{sec:setup}).

\item We identify the sources of performance differences among bi-encoder, cross-encoder, and generative matchers, including the role of pretrained representation geometry, joint pair encoding, prediction formulation, model scaling, and distribution shift (\S~\ref{sec:three-layer}).

\item We extend the comparison beyond within-dataset effectiveness through cross-dataset transfer experiments and a systematic analysis of training time, memory consumption, inference throughput, latency, and representation reuse (\S~\ref{sec:cross}).

\item Based on these findings, we provide practical guidance for EM practitioners, research directions for the EM community, and design principles for future EM benchmarks (\S~\ref{sec:recommendations}).

\item We release our code, training scripts, and evaluation data at \url{https://github.com/Jantory/llm-trained-matcher}.

\end{itemize}
\section{Background and Problem Setup}
\label{sec:background}

Entity matching aims to identify records that refer to the same real-world entity across one or more data sources. A typical EM pipeline consists of two stages. Candidate generation first retrieves record pairs that are potentially matching (typically via a cheap blocking function ~\cite{chantaranimi2023evaluation, brinkmann2024sc, thirumuruganathan2021deep, wang2024neural}), after which a matcher determines whether each candidate pair refers to the same entity. This work focuses on language model-based pairwise matchers~\cite{li2020deep, li2023effective, peeters2025entity} and studies how their behavior varies across matcher architectures, pretraining objectives, and model sizes.

\header{Pairwise entity matching} Let $\mathcal{D}$ denote an EM dataset. Each example in $\mathcal{D}$ is a labeled candidate pair $(r_i, r_j, y_{ij})$, where $r_i$ and $r_j$ are records and $y_{ij} \in {0,1}$ indicates whether the two records refer to the same real-world entity. A record may consist of structured attribute-value pairs, free-form text, or a mixture of both. For structured records, a record is noted as $r = \{(a_1, v_1), \ldots, (a_m, v_m)\}$, where $a_k$ is an attribute name and $v_k$ is the corresponding value.
The matching problem is to learn a scoring function
\[
s_\theta(r_i, r_j) \in [0,1],
\]
where $s_\theta(r_i, r_j)$ estimates the probability that $r_i$ and $r_j$ match. A binary prediction is obtained by thresholding the score:
\[
\hat{y}_{ij} = \mathbb{I}[s_\theta(r_i, r_j) \geq \tau],
\]
where $\tau$ is a decision threshold selected on validation data. 

\header{From records to language model inputs} Structured entity records must be converted into textual inputs before being processed by a large language model. Let $\phi$ denote an input transformation that maps a record $r$ to a sequence $\phi(r)$. Prior work~\cite{hegselmann2023tabllm, xing2025table} shows that this transformation can use different textual templates to linearize the record, with \texttt{attribute: value} being a commonly used choice.
A candidate pair can be represented in two ways. The records can be serialized independently as $\phi(r_i)$ and $\phi(r_j)$, or they can be serialized jointly as
\[
\phi(r_i, r_j) = [\texttt{Record A}: \phi(r_i);   \texttt{ Record B}: \phi(r_j)].
\]
The choice between independent and joint serialization determines where record interaction occurs. This leads to the three matcher architectures studied in this paper.

\subsection{Three LLM-based Matcher Architectures} As illustrated in~\Cref{fig:paradigms}, an LLM can be adapted for EM through three architectures: bi-encoder, cross-encoder, and generative matching. Each architecture can use the same model but differs in how records are encoded and how the matching decision is produced.

\headerul{Bi-encoder matcher} A bi-encoder encodes each record independently. Given records \(r_i\) and \(r_j\), their representations are computed as
\[
h_i = \mathcal{B}_\theta(\phi(r_i)), \qquad h_j = \mathcal{B}_\theta(\phi(r_j)),
\]
where \(\mathcal{B}_\theta\) denotes an LLM-based encoder equipped with a pooling function over the transformer’s hidden states. Unlike encoders based on bidirectional PLMs using the \texttt{CLS} token~\cite{reimers2019sentence}, LLM-based encoders commonly use the hidden state of the final input token \texttt{EOS} as the record representation~\cite{wang2024improving}. Under causal attention~\cite{brown2020language}, this token can attend to all preceding tokens and thus aggregate information from the entire serialized input. The match score is then computed from these two representations:
\[
\mathcal{M}^{\text{bi}}(r_i, r_j) = f(h_i, h_j),
\]
where $f(.)$ may be a similarity function or a learned classifier over the two embeddings. Because $h_i$ and $h_j$ are computed independently, record embeddings can be precomputed and reused across candidate pairs. This makes bi-encoders attractive for efficient retrieval, blocking (candidate generation)~\cite{brinkmann2024sc, zeakis2025depth}, and matching~\cite{karapiperis2026impact}. 

\headerul{Cross-encoder matcher} A cross-encoder matcher encodes the two records jointly~\cite{li2020deep, tu2023unicorn}. The model receives the serialized pair $\phi(r_i, r_j)$ and directly predicts a match score:
\[
\mathcal{M}^{\text{cross}}(r_i, r_j) = \mathcal{C}_\theta(\phi(r_i, r_j)),
\]
where \(\mathcal{C}_\theta\) denotes an LLM-based classifier consisting of a transformer encoder followed by a classification head.
Because both records are present throughout encoding, cross-encoders are therefore expressive pairwise classifiers. However, they are computationally expensive because every entity pair must be processed separately, and record representations cannot be reused across pairs.

\headerul{Generative matcher} A generative matcher retains the language-modeling head of the LLM and formulates matching as conditional generation. Given the serialized pair $\phi(r_i, r_j)$, the model generates a short textual response
\[
z = (z_1, \ldots, z_T) \sim p_\theta(\cdot \mid \phi(r_i, r_j)).
\]
The generated response is then mapped to a binary match decision by an output parser:
\[
\hat{y}_{ij}^{\text{gen}} = \pi(z),
\]
where $\pi(z) \in \{0,1\}$. Generative matchers typically produce a textual decision rather than an explicit real-valued match score. Although a pseudo-score can be derived from the model probabilities or logits associated with predefined output tokens, such as \texttt{yes} and \texttt{no}, our evaluation uses the generated textual decision directly. The generative approach aligns naturally with instruction following language models. However, this architecture also has various disadvantages: auto-regressive generation introduces decoding overhead, and prediction reliability may depend on whether the generated response conforms to the expected output format.
\section{Experimental Design}
\label{sec:setup}

Our experimental design is structured to separate the effects of matching architecture, model variant (representing different pretraining objectives), and model size. The core evaluation follows a full factorial design over: three architectures, three model variants, three model sizes, nine datasets, and five repetitions with different random seeds, for a total of $
3 \times 3 \times 3 \times 9 \times 5 = 1,215$ fine-tuning runs. Each trained model is evaluated under two settings. In within-dataset evaluation, the training and test partitions are drawn from the same dataset. In cross-dataset evaluation, a model trained on one dataset is evaluated on a different dataset.

All experiments were conducted on a Linux-based \seqsplit{high-performance} computing cluster
. Each run had a single NVIDIA H100 SXM5 GPU with 94 GiB of memory, 16 AMD EPYC 9334 CPU cores, and 192 GiB of system memory available. Collectively, the training runs consumed more than 850 GPU-hours.

\subsection{Datasets}
\label{sec:datasets}

We evaluate the matchers on nine ER datasets selected from three widely used benchmarks: Magellan~\cite{mudgal2018deep}, MaChAmp~\cite{cikm21machamp}, and WDC Products~\cite{peeters2023wdc}. The datasets cover product, software, publication, book, and movie matching, and span both schema-aligned structured data and more heterogeneous settings with semi-structured data. Table~\ref{tab:datasets} summarizes their main characteristics.

\begin{table}[t]
\centering
\footnotesize
\caption{Nine evaluation datasets. CC denotes corner cases. Split reports the sizes of the training, validation, and test partitions. Pos. denotes the overall positive-pair rate. \#Attr. and Length report the rounded average numbers of attributes and tokenized record lengths for the two sources, respectively.}
\label{tab:datasets}
\resizebox{\columnwidth}{!}{%
\begin{tabular}{lllcccc}
\toprule
\textbf{ID} & \textbf{Dataset} & \textbf{Type} & \textbf{\#Attr.} & \textbf{Split Size} & \textbf{Pos.} & \textbf{Length} \\
\midrule
\texttt{ab}  & Abt-Buy~\cite{mudgal2018deep}          & struct.        & 3/3   & 5,743/1,916/1,916  & 10.7\% & 90/43 \\
\texttt{ag}  & Amazon-Google~\cite{mudgal2018deep}    & struct.          & 3/3   & 6,874/2,293/2,293  & 10.2\% & 27/33 \\
\texttt{ds}  & DBLP-Scholar~\cite{mudgal2018deep}     & struct.          & 4/4   & 17,223/5,742/5,742 & 18.6\% & 41/42 \\
\texttt{wa}  & Walmart-Amazon~\cite{mudgal2018deep}   & struct.          & 5/5   & 6,144/2,049/2,049  & 9.4\%  & 49/54 \\
\texttt{rt}  & Rel-Text~\cite{cikm21machamp}         & heter.       & 1/4   & 7,417/2,473/2,473  & 18.0\% & 163/51 \\
\texttt{sh}  & Semi-Heter~\cite{cikm21machamp}       & heter.       & 12/12 & 1,240/414/414      & 38.2\% & 137/110 \\
\texttt{sr}  & Semi-Rel~\cite{cikm21machamp}         & heter.      & 8/14  & 1,309/437/437      & 41.6\% & 120/146 \\
\texttt{w20} & WDC (20\% CC)~\cite{peeters2023wdc}    & struct.          & 5/5   & 6,000/3,500/4,500  & 17.9\% & 147/146 \\
\texttt{w80} & WDC (80\% CC)~\cite{peeters2023wdc}    & struct. & 5/5   & 6,000/3,500/4,500  & 17.9\% & 149/149 \\
\bottomrule
\end{tabular}
}
\end{table}

These datasets are deliberately selected to avoid overly simple matching scenarios and to cover a diverse range of matching settings. From Magellan, we exclude datasets that are nearly linearly separable~\cite{papadakis2024critical} or contain very small candidate sets that have less than 2,000 entity pairs. Because MaChAmp datasets were constructed by reusing Magellan-style sources, we retain only datasets whose underlying records are disjoint from those selected from Magellan and on which existing methods have not achieved near-saturated performance~\cite{cikm21machamp, steiner2025fine}. 
From WDC Products, we use two challenging variants containing different proportions of corner cases, defined as hard negatives with high syntactic similarity. We choose the most challenging entity-disjoint splits, such that entities in the test partition do not appear in the training or validation partitions. Overall, the selected datasets vary substantially in labeled-set size, class balance, schema alignment, attribute structure, and serialized record length, providing a diverse testbed for comparing matcher behavior. A common concern in LLM-based research is that public benchmark data may have appeared in the LLM pretraining corpora, which potentially inflates absolute performance through memorization~\cite{schwarzschild2024rethinking}. However, our design choice of focusing on relative comparisons between the same model family and data splits across matcher architectures reduces the impact of potential data contamination on our results.

\subsection{Models and Implementation Details}
\label{sec:models-implementation}

In this study, we restrict our main experiments to decoder-only language models~\cite{brown2020language}, commonly referred to as large language models (LLMs), because generative matchers require a decoder-based backbone. All the training and evaluation experiments are implemented with the Hugging Face \seqsplit{\texttt{transformers}} library~\cite{jain2022hugging}, a widely adopted framework for developing and deploying large language models.

\headerul{Models} Our experiments use the Qwen3 model family~\cite{yang2025qwen3} at three sizes, 0.6B, 4B, and 8B parameters, and model variants corresponding to three pretraining objectives: base, instruction-tuned, and embedding-oriented (as summarized in Table~\ref{tab:model-checkpoints}). The base variant is pretrained on a large-scale corpus using a next-token prediction objective. The instruction-tuned and embedding variants are obtained through additional post-training of the base models for instruction following and text representation, respectively.

\begin{table}[t]
\centering
\caption{Model variants with their model sizes and corresponding hidden dimensions.}
\label{tab:model-checkpoints}
\resizebox{\columnwidth}{!}{%
\begin{tabular}{llc}
\toprule
\textbf{HuggingFace model ID} & \textbf{Training objective} & \textbf{Hidden dim.} \\
\midrule
\texttt{Qwen/Qwen3-\{0.6,4,8\}B-Base} & Next-token prediction & \{1024,2560,4096\} \\
\texttt{Qwen/Qwen3-\{0.6,4,8\}B} & Instruction following & \{1024,2560,4096\} \\
\texttt{Qwen/Qwen3-Embedding-\{0.6,4,8\}B} & Text embedding & \{1024,2560,4096\} \\
\texttt{FacebookAI/roberta-base} & Masked language modeling & 768 \\
\texttt{sentence-transformers/stsb-roberta-base-v2} & Text embedding & 768 \\

\bottomrule
\end{tabular}
}
\end{table}

\headerul{Input serialization} We serialize every record for each dataset as
\[
\phi(r)
=
\texttt{[field\_1] value\_1 ; \ldots ; [field\_k] value\_k}.
\]
For structured datasets with aligned schemas, missing values are retained and replaced with the placeholder \texttt{unknown}; for semi-structured datasets with sparse and heterogeneous fields, missing fields are omitted, following the preprocessing convention of the MaChAmp benchmark~\cite{cikm21machamp}.
Afterwards, the bi-encoder serializes record independently by combining with a task instruction as suggested by ~\cite{yang2025qwen3}. For the cross-encoder and generative matcher, the two serialized records are concatenated into a single input, with the task instruction and the prefixes \texttt{Entity A:} and \texttt{Entity B:} distinguishing the two records. This design follows the same principle as prior work~\cite{zhang2024directions, peeters2025entity, steiner2025fine}.

\begin{tcolorbox}[
    boxrule=0.8pt,
    breakable,
    left=3pt,
    right=3pt,
    top=3pt,
    bottom=3pt]
\noindent\textbf{Task instructions--}

\smallskip
\small
\noindent\textbf{Bi-encoder:} Represent the entity for entity matching.

\smallskip
\noindent\textbf{Cross-encoder:} Determine whether the following two entities refer to the same real-world entity.

\smallskip
\noindent\textbf{Generative matcher:} Determine whether the following two entities refer to the same real-world entity. Answer yes or no.
\end{tcolorbox}

\noindent A common \texttt{max\_len = 512} configuration is applied across all experiments. For bi-encoders, each record is tokenized and truncated independently to at most \texttt{max\_len} tokens. For cross-encoders and generative matchers, each record is truncated to at most \texttt{max\_len}/2 tokens before the paired input is constructed with task instructions and record delimiters. This allocation prevents either record from consuming the full pairwise input budget in rare cases.

\headerul{Bi-encoders} The bi-encoder instantiates the shared encoder \(\mathcal{B}_\theta\) using Hugging Face’s \texttt{AutoModel}, which loads the Transformer backbone without the language-modeling head. Following the pooling strategy defined in~\S\ref{sec:background}, the hidden state of the final input token \texttt{EOS} is used as the record representation.  Let \(h_i\) and \(h_j\) be resulted representations and let $d_{ij} = 1 - \cos(h_i,h_j)$. For label \(y_{ij}\in\{0,1\}\), we optimize the following contrastive loss during training $\mathcal{L}_{\mathrm{bi}}(i,j) = y_{ij} d_{ij}^{2} + (1-y_{ij}) \max(0,m-d_{ij})^{2}$, with margin $m=0.3$. This objective follows the margin-based contrastive learning paradigm commonly used for bi-encoder representation learning~\cite{ruan2025fine, arora2024contrastive}. We also experimented with an InfoNCE objective~\cite{brinkmann2024sc} commonly used for representation-based blocking. However, because the objective relies on in-batch negatives, its performance was sensitive to batch composition and exhibited unstable training behavior and worse performance. 

\headerul{Cross-encoders} The cross-encoder instantiates the classifier \(\mathcal{C}_\theta\) using Hugging Face’s \texttt{\seqsplit{AutoModelForSequenceClassification}}, which loads the Transformer backbone together with a task-specific sequence-classification head. The model is trained as a binary classifier over candidate pairs using cross-entropy loss over the two class logits. At inference time, we use the softmax probability of the positive class as the match score for threshold-based metrics.

\headerul{Generative matchers} The generative matcher instantiates \(p_\theta\) using Hugging Face’s \texttt{\seqsplit{AutoModelForCausalLM}}, which loads the transformer backbone together with the causal language-modeling head used to generate the match decision auto-regressively. The target output is \texttt{yes} for a matching pair and \texttt{no} for a non-matching pair. We optimize the standard causal language modeling loss only on the response tokens, masking the prompt tokens from the loss. At inference time, the model generates up to five tokens. We strip leading and trailing whitespace from the generated text and extract its first whitespace-delimited word, which is mapped to a positive prediction if it is \texttt{yes} and a negative prediction if it is \texttt{no} using case-insensitive matching. Any other output is considered invalid.

\subsection{Training Protocol and Evaluation Metrics}
\label{sec:training-eval}

All model configurations are fine-tuned using the Hugging Face \texttt{Trainer} API under a standardized training budget and five random seeds. Each run uses a batch size of 64, a maximum of 400 optimization steps, validation runs every 20 steps, early stopping with a patience of 5 validation rounds, a warm up ratio of 0.05, and gradient checkpointing is enabled to optimize memory consumption. To standardize the training budget across datasets, the training and validation splits are randomly capped at 8,000 and 2,000 labeled pairs, respectively, whereas each test set is evaluated in full.

LoRA adapters are used across all configurations to reduce memory usage. This choice is supported by prior work showing that LoRA achieves performance comparable to full fine-tuning for entity matching while updating only a small fraction of the model parameters~\cite{vos2022towards,zhang2024directions}. We set the rank to 8, the scaling factor to 16, and the dropout rate to 0.05, and apply adapters to all linear layers following the default recommendation of the PEFT library ~\cite{mangrulkar2022peft}. F1 (in \%) serves as the primary metric, while precision, recall, and AUROC are also used for diagnostic analyses where appropriate.

Before conducting the full experiments, the learning rate was tuned separately for each combination of matching architecture, model variant, and model size over five candidate values using an auxiliary tuning dataset. The dataset was constructed by sampling a total of 800 labeled pairs exclusively from the training partitions of all used datasets. These pairs were divided into 500 training, 200 validation, and 100 held-out tuning pairs. For each configuration, the learning rate achieving the highest F1 score on the held-out tuning split over five random seed runs was used for all dataset-specific runs. No examples from the original validation or test partitions were used for hyperparameter tuning.

\subsection{Baselines}
\label{sec:baselines}

We include three representative language model-based baselines aligned with the matching architectures studied in this work: \texttt{Sentence -BERT}~\cite{reimers2019sentence, zeakis2025depth} for bi-encoder matching, \texttt{Ditto}~\cite{li2020deep} for cross-encoder matching, and \texttt{GPT-4.1-mini}\footnote{\url{https://developers.openai.com/api/docs/models/gpt-4.1-mini}} for generative matching. Together, these baselines provide complementary reference points for assessing the effects of model size, architecture-specific training, and task-specific fine-tuning.

\headerl{Sentence-BERT} We adapt our bi-encoder implementation to fine-tune \texttt{\seqsplit{sentence-transformers/stsb-roberta-base-v2}}. The two records are encoded independently with a shared RoBERTa encoder~\cite{liu2019roberta}, and their representations are obtained by mean pooling over the final hidden states. All other settings follow the main bi-encoder implementation, except for the learning rate. Differing from the Qwen3-based encoders, it may not share the same suitable learning-rate range. Thus, the recommended Sentence-BERT learning rate~\cite{reimers2019sentence} is used to avoid disadvantaging the baseline through an inappropriate optimization setting.

\headerl{Ditto} We use the official Ditto implementation~\cite{li2020deep} with its proposed optimization techniques enabled, including domain knowledge injection, summarization, and data augmentation. Ditto is built on RoBERTa~\cite{liu2019roberta} and is a widely adopted supervised transformer baseline for entity matching. We train and evaluate Ditto on the same dataset splits as used in our main experiments. 

\headerl{GPT-4.1-mini} GPT-4.1-mini serves as the generative baseline. It was selected as a cost-efficient successor to GPT-4o-mini, which has been examined in prior entity-matching studies~\cite{zhang2025deep, peeters2025entity}, while offering stronger performance on general model benchmarks. It therefore provides a more capable contemporary reference point for API-based generative matching. The model is evaluated in a zero-shot setting. Each entity pair is serialized using the same pairwise input format as the fine-tuned generative matchers and accompanied by an instruction to return a binary match decision.


\section{Analysis Overview and Main Findings}
\label{sec:analysis-overview}

This section provides a roadmap for the empirical analyses that follow. It introduces the main dimensions along which matching architectures, model variants, and model sizes are compared, summarizes the key findings, and directs the reader to the subsequent sections for detailed results and supporting evidence.

\subsection{Analysis Roadmap}

The analysis begins with controlled within-dataset comparisons in~\S\ref{sec:three-layer}. We first examine how performance depends on initialization with a given model variant. Based on these results, we select the best variant for each matcher architecture and use it in the subsequent analyses. We then evaluate the benefits of jointly encoding entity pairs and identify the conditions under which cross-encoder and generative matcher decisions diverge. The observed divergence is driven by two particular datasets, motivating further analysis. By studying the original data partitions and constructing rebalanced variants, we trace the performance differences to unseen subtle schema differences and class-conditional feature distribution shift between the training and evaluation sets. 
\S\ref{sec:cross} extends the evaluation to transfer across datasets and returns to the comparison among the three matcher architectures. Specifically, it measures how much target-specific performance is retained when the training supervision is obtained from a difference source dataset, thereby revealing the extent to which fine-tuned behavior remains tied to the source distribution.
Finally, \S\ref{sec:cost} relates predictive performance to training time, memory consumption, and inference throughput. This breakdown identifies the deployment conditions under which the computational properties of each architecture outweigh and justify their effectiveness differences.

\subsection{Main Findings}
We summarize the following six key findings, which are detailed in the coming sections.

\begin{enumerate}[leftmargin=*]
\item Model variants that correspond to different pretraining objectives matter primarily for bi-encoders. Embedding-oriented variants provide substantially better initialization for bi-encoders fine-tuning than other variants. Their pretrained representation space already separates matching from non-matching pairs and can predict downstream performance (\S\ref{sec:three-layer:representation}).  

\item Joint pair encoding provides a persistent advantage over independent encoding. Although increasing the model size improves bi-encoder performance, it does not eliminate the gap relative to cross-encoders, even at the 8B scale (\S\ref{sec:three-layer:interaction}). 

\item Generative matching offers a conditional but not consistent advantage. Cross-encoder and generative matchers perform similarly on most datasets. The generative advantage is concentrated on cases with pronounced differences between training and evaluation distributions (\S\ref{sec:three-layer:generalization}). 

\item Different distribution shifts produce distinct failure mode. Unseen schema combinations primarily reduce cross-encoder recall, whereas an increased prevalence of same-title negative pairs causes larger model to over-predict matches and lose precision (\S\ref{sec:three-layer:generalization}).

\item Matching architectures differ remarkably in cross-dataset transferability. Bi-encoders retain the least target-specific performance under cross-dataset evaluation, while generative matchers achieve the strongest overall retention. Besides, the transferability also varies by source and target dataset (\S\ref{sec:cross}).

\item Architecture choice depends on the deployment conditions. Cross-encoders provide the strongest effectiveness-throughput balance for pairwise matching. Generative matchers are preferable in  settings with distribution shift, but incur a high memory usage. Bi-encoders become competitive when the representations can be cached and reused along with blocking (\S\ref{sec:cost}). 
\end{enumerate}

\section{Disentangling the Effects of Matcher Design}
\label{sec:three-layer}

This section first examines how bi-encoder performance depends on the pretrained representation space (induced by the choice of model variant), then assesses what direct interaction adds by comparing bi-encoders with cross-encoders. Finally, it compares cross-encoder and generative matchers to identify the datasets and conditions where their decisions diverge.

\subsection{Effect of Pre-training Objective}
\label{sec:three-layer:representation}

\begin{table}[t]
\centering
\small
\caption{Effect of pretraining objective for matching architectures. F1 is macro-averaged over datasets and model sizes. Wins counts datasets where the variant obtains the best F1.}
\label{tab:rq1_variant}
\resizebox{\columnwidth}{!}{
\begin{tabular}{l cc cc cc cc}
\toprule
& \multicolumn{2}{c}{\textbf{Instruct}}
& \multicolumn{2}{c}{\textbf{Base}}
& \multicolumn{2}{c}{\textbf{Embedding}}
& \multicolumn{2}{c}{\textbf{Best vs.\ 2nd}}\\
\cmidrule(lr){2-3}
\cmidrule(lr){4-5}
\cmidrule(lr){6-7}
\cmidrule(lr){8-9}
\textbf{Architecture} & F1 & wins & F1 & wins & F1 & wins & \(\Delta\) & \(p\) \\
\midrule
Bi-encoder & 53.7 & 0 & 57.4 & 0 & \underline{77.9} & \underline{9} & \(+20.3\) & 0.004 \\
Cross-encoder & 82.5 & 4 & \underline{83.7} & \underline{5} & 81.7 & 0 & \(+1.2\) & 0.652 \\
Generative & 85.9 & 0 & 86.3 & \underline{7} & \underline{86.4} & 2 & \(<0.1\) & 0.570 \\
\bottomrule
\end{tabular}
}
\end{table}

The bi-encoder architecture encodes two records independently before computing the matching score. Its performance therefore depends strongly on how pretraining shapes the resulting representation space.

We compare model variants corresponding to different pretraining objectives. Table~\ref{tab:rq1_variant} reports macro-averaged test F1 across datasets and model sizes. For the wins count, we first select the best F1 over model sizes for each dataset and checkpoint variant, and then count the datasets on which each variant performs best. We also report \(\Delta\), defined as the difference between the best and second-best variants in macro-averaged F1, together with a two-sided Wilcoxon \(p\)-value computed over the nine per-dataset means.

For the bi-encoder, the embedding variant achieves an F1 score of 77.9, ranks first on all nine datasets, and exceeds the second-best variant by 20.3 points; the difference is statistically significant (\(p=0.004\)). By contrast, the choice of model variant has a much smaller effect on cross-encoder and generative matchers. Surprisingly, even an embedding variant can be fine-tuned into an effective generative matcher. This suggests that fine-tuning can largely attenuate the effect of the pretraining objective for cross- and generative matchers.

\begin{figure}[t]
\centering
\includegraphics[width=\columnwidth]{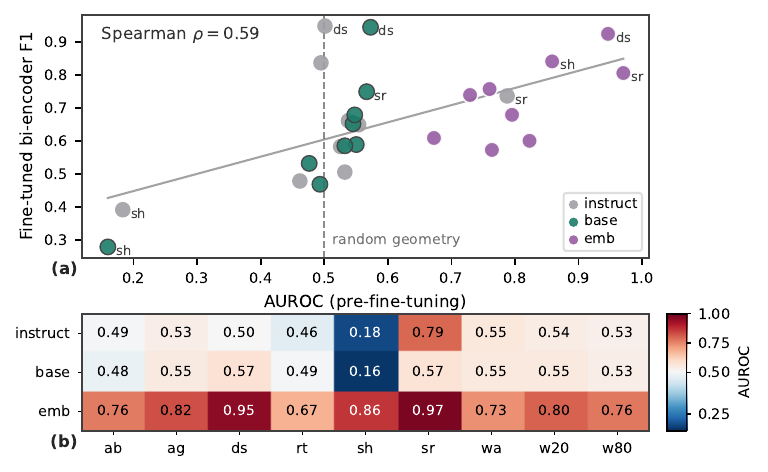}
\caption{Pretrained representation geometry predicts bi-encoder performance at 0.6B. (a) Zero-shot cosine-similarity AUROC correlates with fine-tuned F1 across dataset--checkpoint pairs (\(\rho=0.59\)). (b) The embedding variant achieves the highest zero-shot AUROC on all nine datasets.}
\label{fig:geometry-probe}
\end{figure}

The dominance of the embedding variant suggests that the performance gap originates from the pretrained representation space. The contrastive pretraining of embedding variants provides a more favorably shaped space than base or instruction-tuned variants, whose hidden states are not explicitly optimized for representation-level comparison. We validate this hypothesis by encoding test records with each 0.6B variant before EM fine-tuning and computing the AUROC of pairwise cosine similarity. AUROC is a suitable indicator here because it measures whether matching pairs are ranked above non-matching pairs
independently of any threshold calibration.
Figure~\ref{fig:geometry-probe} shows that the embedding variant provides a strong initial matching signal on all nine datasets, whereas the base and instruction-tuned variant are close to random on most datasets. On \texttt{Semi-Heter}, their AUROC falls below 0.5, indicating that non-matches are ranked above matches, while the embedding variant reaches 0.858. Across the 27 dataset--variant pairs, zero-shot AUROC correlates with fine-tuned bi-encoder F1 (\(\rho=0.59\)). This correlation indicates that the pretrained representation space is already highly informative of downstream performance.

To exclude that LoRA causes this dependence on variant selection, we fully fine-tune the 0.6B base and instruction-tuned bi-encoders on three datasets where these variants perform poorly: \texttt{ABT-Buy}, \texttt{Semi-Heter}, and \texttt{Semi-Rel}. Full fine-tuning closes the gap on \texttt{ABT-Buy}, reaching F1 scores of 75.4 and 75.0 compared with 75.7 for the embedding variant with LoRA. It remains substantially worse on \texttt{Semi-Heter} (32.9 and 41.1 versus 84.0) and \texttt{Semi-Rel} (66.8 and 73.7 versus 80.5). This indicates that the observed dependence on variant selection cannot be attributed to low-rank adaptation.

\subsection{Effect of Joint Pair Encoding}
\label{sec:three-layer:interaction}

\begin{table*}[!t]
\centering
\small
\begin{minipage}[t]{0.67\textwidth}
\centering
\caption{Per-dataset test F1 scores and scaling gains across matching architectures. The left (a) reports test F1 (mean $\pm$ standard deviation over five seeds) at each model size, using the best model variant for each architecture together with its corresponding baseline. The best score in each row is underlined. The right (b) illustrates dataset-level F1 gains between adjacent model sizes for each matching architecture.}
\label{tab:per_dataset_scale}
\vspace{0.35em}
\setlength{\tabcolsep}{3pt}
\resizebox{\textwidth}{!}{%
\begin{tabular}{l cccccccccccc}
\toprule
 & \multicolumn{4}{c}{\textbf{Bi-encoder}} & \multicolumn{4}{c}{\textbf{Cross-encoder}} & \multicolumn{4}{c}{\textbf{Generative}} \\
\cmidrule(lr){2-5}\cmidrule(lr){6-9}\cmidrule(lr){10-13}
Dataset & SBERT & 0.6B & 4B & 8B & Ditto & 0.6B & 4B & 8B & GPT & 0.6B & 4B & 8B \\
\midrule
\texttt{ab} & $87.5_{\pm0.2}$ & $75.7_{\pm0.4}$ & $88.0_{\pm0.3}$ & $91.1_{\pm0.9}$ & $84.7_{\pm3.5}$ & $93.0_{\pm1.3}$ & $94.0_{\pm0.5}$ & $\underline{94.9}_{\pm0.4}$ & $87.2$ & $93.4_{\pm0.6}$ & $93.5_{\pm0.8}$ & $94.4_{\pm0.5}$ \\
\texttt{ag} & $73.7_{\pm0.6}$ & $60.0_{\pm0.5}$ & $72.0_{\pm0.6}$ & $76.3_{\pm0.5}$ & $72.2_{\pm1.2}$ & $79.1_{\pm1.3}$ & $78.7_{\pm3.6}$ & $78.6_{\pm2.8}$ & $64.6$ & $77.9_{\pm2.7}$ & $78.1_{\pm1.9}$ & $\underline{80.6}_{\pm1.6}$ \\
\texttt{ds} & $93.3_{\pm0.1}$ & $92.4_{\pm0.4}$ & $93.9_{\pm0.9}$ & $95.3_{\pm0.4}$ & $94.9_{\pm0.3}$ & $95.0_{\pm0.2}$ & $95.1_{\pm0.3}$ & $95.2_{\pm0.6}$ & $87.4$ & $\underline{95.5}_{\pm0.6}$ & $95.3_{\pm0.4}$ & $95.0_{\pm0.7}$ \\
\texttt{rt} & $61.7_{\pm0.8}$ & $60.9_{\pm0.6}$ & $67.5_{\pm0.6}$ & $72.0_{\pm0.7}$ & $38.5_{\pm17.6}$ & $73.2_{\pm1.1}$ & $75.7_{\pm1.1}$ & $\underline{75.8}_{\pm1.2}$ & $69.5$ & $73.4_{\pm0.5}$ & $75.3_{\pm0.8}$ & $74.8_{\pm0.8}$ \\
\texttt{sh} & $55.3_{\pm0.5}$ & $84.0_{\pm1.0}$ & $86.7_{\pm0.7}$ & $86.0_{\pm1.7}$ & $56.3_{\pm3.8}$ & $58.7_{\pm0.6}$ & $62.8_{\pm16.9}$ & $65.1_{\pm14.4}$ & $\underline{94.0}$ & $73.2_{\pm14.0}$ & $88.8_{\pm3.9}$ & $91.5_{\pm2.9}$ \\
\texttt{sr} & $82.2_{\pm0.7}$ & $80.5_{\pm0.6}$ & $79.9_{\pm0.4}$ & $85.6_{\pm1.7}$ & $92.6_{\pm3.6}$ & $\underline{97.6}_{\pm0.2}$ & $89.5_{\pm4.7}$ & $85.0_{\pm11.4}$ & $87.2$ & $94.0_{\pm5.2}$ & $88.3_{\pm3.1}$ & $90.0_{\pm5.8}$ \\
\texttt{wa} & $77.6_{\pm0.6}$ & $73.9_{\pm0.6}$ & $82.0_{\pm1.2}$ & $81.9_{\pm0.5}$ & $85.5_{\pm1.6}$ & $88.5_{\pm0.8}$ & $\underline{91.6}_{\pm0.4}$ & $90.7_{\pm0.4}$ & $78.2$ & $89.8_{\pm0.7}$ & $90.8_{\pm0.5}$ & $90.6_{\pm0.6}$ \\
\texttt{w20} & $67.9_{\pm0.5}$ & $67.9_{\pm0.6}$ & $75.0_{\pm0.9}$ & $76.2_{\pm0.3}$ & $76.4_{\pm1.0}$ & $84.8_{\pm1.0}$ & $87.8_{\pm0.8}$ & $88.7_{\pm1.0}$ & $87.3$ & $83.8_{\pm0.4}$ & $88.4_{\pm0.2}$ & $\underline{88.9}_{\pm0.3}$ \\
\texttt{w80} & $56.6_{\pm1.0}$ & $57.3_{\pm0.2}$ & $66.3_{\pm1.1}$ & $69.4_{\pm0.9}$ & $57.9_{\pm21.4}$ & $77.0_{\pm0.5}$ & $81.7_{\pm0.8}$ & $83.2_{\pm0.5}$ & $82.2$ & $78.9_{\pm0.4}$ & $82.6_{\pm0.4}$ & $\underline{83.9}_{\pm1.3}$ \\
\midrule
\textbf{Average} & $72.9_{\pm0.6}$ & $72.5_{\pm0.5}$ & $79.0_{\pm0.8}$ & $81.5_{\pm0.8}$ & $73.2_{\pm6.0}$ & $83.0_{\pm0.8}$ & $84.1_{\pm3.2}$ & $84.1_{\pm3.6}$ & $82.0$ & $84.4_{\pm2.8}$ & $86.8_{\pm1.3}$ & $\underline{87.7}_{\pm1.6}$ \\
\bottomrule
\end{tabular}%
}
\end{minipage}%
\hfill%
\begin{minipage}[t]{0.32\textwidth}
\raggedright
\vspace{0pt}
\centering
\includegraphics[width=\textwidth]{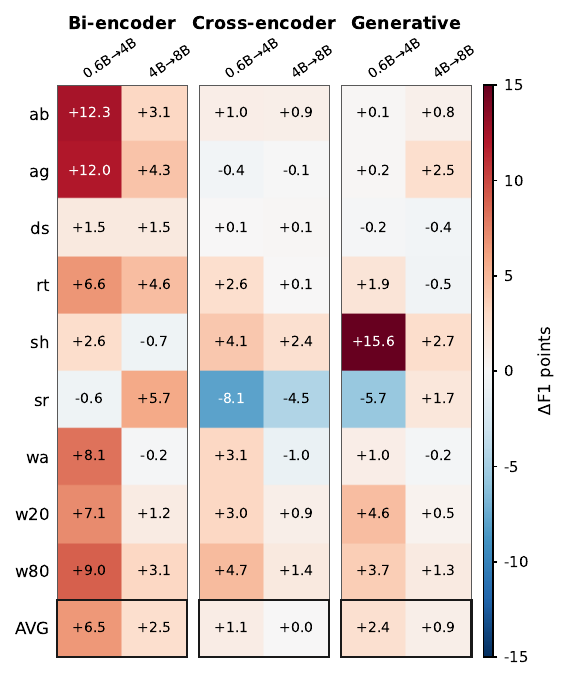}
\end{minipage}
\end{table*}

We next investigate the effect of joint pair encoding by primarily comparing the bi-encoder and the cross-encoder. Unless otherwise stated, the remaining analyses use the best-performing model variant for each matching architecture: the embedding variant for bi-encoders and the base variant for cross-encoders and generative matchers, as indicated by Table~\ref{tab:rq1_variant}.

Table~\ref{tab:per_dataset_scale} shows that bi-encoder performance improves consistently with model size, from 72.5 at 0.6B to 79.0 at 4B and 81.5 at 8B. Nevertheless, the bi-encoders remain below cross-encoders at every model size, which achieve average F1 scores of 83.0, 84.1, and 84.1. Model size scaling improves independently encoded representations but does not eliminate the gap introduced by the absence of direct pairwise interaction.

The baseline comparison shows a similar pattern. SBERT and Ditto exhibit the same distinction with a shared RoBERTa backbone: SBERT encodes the two records independently, whereas Ditto jointly encodes the pair. Ditto outperforms SBERT on average. This parallel finding suggests that the cross-encoder advantage is not specific to Qwen3 or a particular variant, but reflects a general benefit of joint pair encoding for EM.

The EM decision often requires fine-grained field level comparisons across records to align abbreviations, interpret conflicting values, etc. A bi-encoder must compress each record into a single representation before observing the other, which limits its ability to capture such local correspondences. A cross-encoder instead allows the records to interact during encoding. Thus, even stronger representations from larger models improve bi-encoder performance, but explicit pairwise interaction remains a distinct source of predictive gain.

\begin{figure}[t]
    \centering
    \includegraphics[width=0.95 \columnwidth]{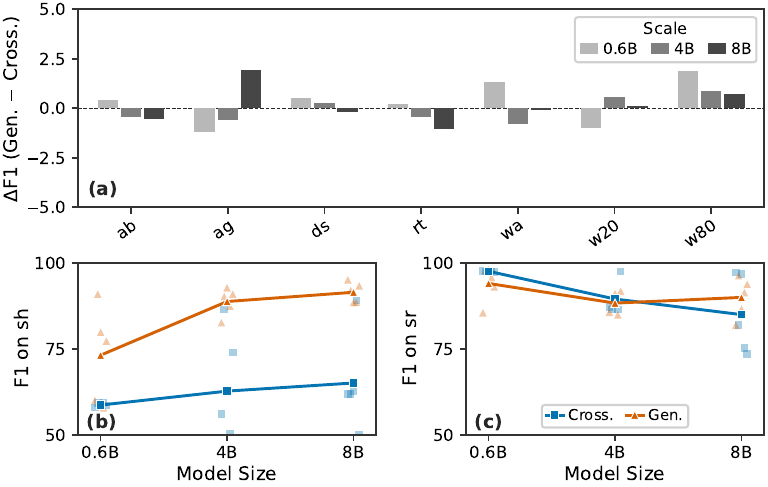}
    \caption{Cross-encoders vs generative matchers. (a) Mean \(\Delta\)F1 on seven datasets where they perform similarly. (b)--(c) Model size-dependent F1 on \texttt{sh} and \texttt{sr}, respectively.}
    \label{fig:cross-generative}
\end{figure}

\subsection{Effect of Matching Prediction Formulation}
\label{sec:three-layer:generalization}

We next examine the differences between cross-encoder and generative matchers. We observe that every fine-tuned generative matcher emits a valid \texttt{yes} or \texttt{no} token at the first decoding step. This behavior allows us to interpret the comparison between the two matching architectures primarily as a comparison between their prediction formulations: a task-specific sequence-classification head for the cross-encoder versus a causal language-modeling head for the generative matcher. The classification head is initialized and trained from scratch, while the language-modeling head is adapted to the matching task during fine-tuning.

Overall, generative matchers achieve higher performance, as shown in Table~\ref{tab:per_dataset_scale}. Cross-encoder F1 increases from 83.0 to 84.1 and then remains at 84.1 across the three model sizes, whereas generative matcher performance rises from 84.4 to 86.8 and 87.7. The same pattern is observed for the two baselines: zero-shot GPT achieves an F1 of 82.0, exceeding Ditto's 73.2 by 8.8 points. 
However, this aggregate advantage is not uniform and concentrated on the \texttt{Semi-Heter} and \texttt{Semi-Rel} datasets. After excluding these two datasets, the macro-averaged F1 scores of cross-encoders are 84.4, 86.4, and 86.7 across the three model sizes, compared with 84.7, 86.3, and 86.9 for generative matchers. Figure~\ref{fig:cross-generative}(a) illustrates that the two architectures perform nearly identically on the remaining seven datasets. By contrast, Table~\ref{tab:per_dataset_scale}(b) and Figure~\ref{fig:cross-generative}(b)--(c) shows pronounced divergences between two architectures on \texttt{Semi-Heter} and \texttt{Semi-Rel}. This motivates a detailed analysis of the observed divergences on \texttt{Semi-Heter} and \texttt{Semi-Rel} to identify when generative matching provides benefits beyond joint pair encoding. We show that these gaps arise under specific train--test distribution shifts. 

\header{Impact of attribute inconsistencies} We begin with \texttt{Semi-Heter}, a dataset of book records collected from two heterogeneous sources. Records from both sources are semi-structured, with each instance stored as a JSON object whose set of attributes vary across records. The original split exhibits a severe train--test mismatch: 82\% of positive test pairs have attribute combinations that are absent from both the training and validation sets. For example, the attribute-set pair \texttt{(title, pub year, pub month, pub day, publisher, authors, edition, isbn13, language, pages)} and \texttt{(Title, publisher, pages, isbn13)} occurs in exactly one positive pair, and that pair appears exclusively in the test set. The model therefore has to identify matching evidence across attribute combinations that were not observed during training.
As shown in Figure~\ref{fig:cross-generative}(b), both architectures benefit from scaling, but the gap between them remains substantial and widens with model size. At 0.6B, the generative matcher achieves an F1 score of 73.2, compared with 58.7 for the cross-encoder. At 8B, their F1 scores increase to 91.5 and 65.1, respectively. A large gap also appears on the two baselines shown in Table~\ref{tab:per_dataset_scale}(a), where zero-shot GPT substantially outperforms Ditto, with F1 scores of 94.0 and 56.3, respectively.
Table~\ref{tab:sh-balance-diagnostics} further reveals that the difference is driven primarily by recall. On the original split, the cross-encoder maintains high precision at every model size, but its recall remains low at 42.6, 48.9, and 51.1. The generative matcher also has relatively low recall at 0.6B, but benefits much more from scaling, reaching a recall of 84.8 at 8B while maintaining high precision.

To assess whether the shift in attribute combinations accounts for this divergence between the two architectures, we construct a rebalanced version of \texttt{Semi-Heter} in which the proportion of positive test pairs with unseen attribute combinations is reduced from 82\% to 17\%. After fine-tuning on the rebalanced dataset, cross-encoder recall rises to approximately 96--97 across all model sizes, and its F1 becomes comparable to that of the generative matcher.
These results indicate that the generative matcher is more robust to the attribute-combination shift, suggesting that it generalizes matching evidence more effectively across unseen schema inconsistencies.

\begin{table}[t]
\centering
\caption{Performance on original and balanced \texttt{sh} across model sizes. Parentheses indicate the portion of unseen attribute combinations of positive pairs in test set.}
\label{tab:sh-balance-diagnostics}
\resizebox{\columnwidth}{!}{%
\begin{tabular}{llccc@{\hspace{0.8em}}ccc}
\toprule
&
& \multicolumn{3}{c}{\textbf{Original} (82\% unseen)}
& \multicolumn{3}{c}{\textbf{Balanced} (17\% unseen)} \\
\cmidrule(lr){3-5}\cmidrule(lr){6-8}
\textbf{Architecture} & \textbf{Model Size} & F1 & P & R & F1 & P & R \\
\midrule
\multirow{3}{*}{\textbf{Cross-encoder}}
& 0.6B & $58.7_{\pm0.6}$  & $94.2_{\pm1.1}$ & $42.6_{\pm0.7}$  & $97.1_{\pm0.5}$ & $97.1_{\pm1.0}$ & $97.1_{\pm1.2}$ \\
& 4B   & $62.8_{\pm16.9}$ & $94.0_{\pm2.4}$ & $48.9_{\pm19.9}$ & $97.3_{\pm0.4}$ & $97.6_{\pm0.8}$ & $97.1_{\pm1.2}$ \\
& 8B   & $65.1_{\pm14.4}$ & $94.7_{\pm1.9}$ & $51.1_{\pm18.0}$ & $96.1_{\pm2.5}$ & $96.0_{\pm3.9}$ & $96.3_{\pm1.8}$ \\
\midrule
\multirow{3}{*}{\textbf{Generative}}
& 0.6B & $73.2_{\pm14.0}$ & $97.2_{\pm1.0}$ & $60.3_{\pm18.1}$ & $96.6_{\pm0.6}$ & $98.0_{\pm0.9}$ & $95.2_{\pm1.5}$ \\
& 4B   & $88.8_{\pm3.9}$  & $97.7_{\pm0.5}$ & $81.5_{\pm6.6}$  & $97.8_{\pm0.5}$ & $98.1_{\pm0.8}$ & $97.5_{\pm0.8}$ \\
& 8B   & $91.5_{\pm2.9}$  & $99.4_{\pm0.6}$ & $84.8_{\pm4.8}$  & $96.9_{\pm1.2}$ & $98.8_{\pm0.8}$ & $95.0_{\pm2.8}$ \\
\bottomrule
\end{tabular}%
}
\end{table}

\header{Impact of class-conditional feature distribution shift} We next consider \texttt{Semi-Rel}, a dataset of movie records collected from two heterogeneous sources, one semi-structured and the other structured. We detect a form of class-conditional feature distribution shift between the training and evaluation data: there is a change in how reliably the agreement on one or more attributes indicates a match. We refer to this as attribute agreement shift in the following. This shift concerns on the movie title: a title match easily predicts a record match in the training and validation sets, but becomes substantially less reliable in the test set. For example, the records \texttt{(title: From Time To Time; director: Julian Fellowes; price: \$12.98; year: 2012; duration: 95 minutes)} and \texttt{(title: From Time to Time; audience\_rating: 59; director: Julian Fellowes; year: 2009; time: 1 hr. 35 min.)} share both title and director but form a negative pair because their release years and durations differ. This example illustrates that although title agreement is a strong matching cue during training, it is not a reliable decision rule in the test distribution. More broadly, negative pairs with identical titles account for only 0.5\% and 0.4\% of the training and validation negatives, respectively, but this rate increases more than 70\(\times\) in the test set. 

Such a shift creates a condition for shortcut learning, in which a model relies on an easily identifiable cue whose predictive value does not remain stable across data partitions~\cite{geirhos2020shortcut}. As shown in Table~\ref{tab:per_dataset_scale}(b) and Figure~\ref{fig:cross-generative}(c), both architectures degrade as model size increases, but the decline is more severe for cross-encoders. Cross-encoder F1 drops from 97.6 at 0.6B to 85.0 at 8B, whereas generative matcher F1 decreases from 94.0 to 90.0 over the same range, with a non-monotonic pattern at 4B.
Such a degradation is driven primarily by precision, as shown in Table~\ref{tab:sr-balance-diagnostics}. For cross-encoders, precision falls from 96.9 at 0.6B to 76.4 at 8B, while recall remains close to 99 across model sizes. Generative matchers exhibit a similar but weaker decline, with precision decreasing from 90.9 at 0.6B to 83.7 at 8B. This pattern is consistent with prior observations that larger LLMs can exhibit stronger reliance on spurious correlations~\cite{yuan2024llms}.

To validate our hypothesis, we construct a rebalanced version of the dataset, in which the proportions of negative pairs with identical titles remain similar in different partitions. After fine-tuning on the rebalanced dataset, the degradation disappears: both architectures achieve F1 scores of approximately 98 across all model sizes, with precision restored to approximately 97--98.
These results indicate that the original degradation is driven by shortcut learning under the attribute-agreement shift, with scaling the model size amplifying reliance on the misleading agreement cue. Both architectures are affected, but the cross-encoder becomes increasingly brittle as model size grows.

\begin{table}[t]
\centering
\caption{Performance on original and rebalanced \texttt{sr} across model sizes. Parentheses report the title-collision rate among negative pairs in train/valid/test partitions.}
\label{tab:sr-balance-diagnostics}
\resizebox{\columnwidth}{!}{%
\begin{tabular}{llccc@{\hspace{0.8em}}ccc}
\toprule
&
& \multicolumn{3}{c}{\textbf{Original} (0.5/0.4/28.7\%)}
& \multicolumn{3}{c}{\textbf{Rebalanced} (6.2/6.2/5.9\%)} \\
\cmidrule(lr){3-5}\cmidrule(lr){6-8}
\textbf{Architecture} & \textbf{Model Size} & F1 & P & R & F1 & P & R \\
\midrule
\multirow{3}{*}{\textbf{Cross-encoder}}
& 0.6B & $97.6_{\pm0.2}$  & $96.9_{\pm0.9}$  & $98.3_{\pm0.9}$ & $98.2_{\pm0.4}$ & $97.7_{\pm0.7}$ & $98.7_{\pm1.1}$ \\
& 4B   & $89.5_{\pm4.7}$  & $81.7_{\pm9.0}$  & $99.6_{\pm1.0}$ & $98.3_{\pm0.3}$ & $97.4_{\pm0.4}$ & $99.2_{\pm0.3}$ \\
& 8B   & $85.0_{\pm11.4}$ & $76.4_{\pm19.0}$ & $98.7_{\pm1.4}$ & $98.1_{\pm0.7}$ & $97.4_{\pm0.5}$ & $98.9_{\pm1.0}$ \\
\midrule
\multirow{3}{*}{\textbf{Generative}}
& 0.6B & $94.0_{\pm5.2}$ & $90.9_{\pm10.1}$ & $98.0_{\pm1.4}$ & $98.4_{\pm0.3}$ & $97.6_{\pm0.3}$ & $99.1_{\pm0.7}$ \\
& 4B   & $88.3_{\pm3.1}$ & $80.7_{\pm5.4}$  & $97.7_{\pm0.5}$ & $98.3_{\pm0.2}$ & $97.4_{\pm0.4}$ & $99.2_{\pm0.5}$ \\
& 8B   & $90.0_{\pm5.8}$ & $83.7_{\pm10.5}$ & $98.0_{\pm1.8}$ & $98.6_{\pm0.3}$ & $98.4_{\pm0.8}$ & $98.9_{\pm0.5}$ \\
\bottomrule
\end{tabular}%
}
\end{table}

\section{Cross-dataset Transferability of Fine-tuned Matchers}
\label{sec:cross}
While the within-dataset experiments characterize how architectural and model-level choices affect matcher behavior, they provide limited evidence about whether the resulting decision rules generalize beyond the source distribution. This section therefore examines whether fine-tuning learns transferable EM behavior or source-specific decision rules. Inspired by prior work~\cite{zhang2025deep, steiner2025fine}, we fine-tune each model on one dataset and evaluate it on the others.

\begin{figure*}[t]
\centering
\includegraphics[width=\textwidth]{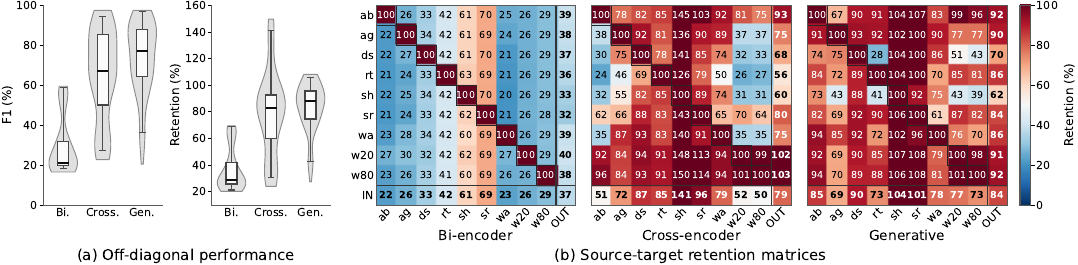}
\caption{Cross-dataset F1 and retention at 8B. Higher retention indicates that fine-tuning on the source preserves more of the target's in-domain performance. In the matrices, rows are fine-tuning sources and columns are evaluation targets.}
\label{fig:cross-dataset-retention}
\end{figure*}

\subsection{Measuring Cross-Dataset Retention}
\label{sec:cross:measure}

We evaluate cross-dataset transfer with the most performative 8B model size. For every source dataset \(s\), we fine-tune a model on \(s\) and evaluate it on every target dataset \(t\), averaging results over five random seeds. For each source--target pair, we define F1 retention as
\[\mathrm{Retention}(s,t)
=
\frac{\mathrm{F1}(\mathrm{train}=s,\mathrm{test}=t)}
{\mathrm{F1}(\mathrm{train}=t,\mathrm{test}=t)} \times 100.
\]
Retention measures how much target-specific performance is preserved when the model is trained on another dataset. A value of 100 indicates that source training matches in-domain fine-tuning on the target, while values above 100 indicate that external training performs better.

\subsection{Architectures Differ in Transferability}
\label{sec:cross:trans}

Figure~\ref{fig:cross-dataset-retention}(a) summarizes cross-dataset transfer performance in terms of both absolute F1 and retention relative to in-domain performance. The three architectures exhibit markedly different levels of transferability. Bi-encoders generalize poorly, achieving a mean off-diagonal F1 of only 30.3, together with mean and median retention scores of 36.9 and 28.8, respectively. Cross-encoders substantially improve cross-dataset performance, increasing the mean F1 to 64.9 and the mean and median retention rates to 79.1 and 83.0. Generative matchers perform best overall, attaining the highest mean off-diagonal F1 of 73.8, as well as mean and median retention rates of 83.6 and 88.4, respectively.

This cross-dataset analysis reveals that the advantages associated with matcher design persist and become more pronounced under distribution shift. Bi-encoder retention is concentrated at low values, suggesting that representations learned on one dataset are rarely reusable across heterogeneous targets. Cross-encoders transfer considerably better, indicating that direct pairwise interaction yields more generalizable matching behavior. Generative matchers retain the largest proportion of target-specific performance, demonstrating the strongest robustness.

We attribute the stronger transfer of generative matchers may to their prediction formulation. Fine-tuning can adapt an existing language-based decision process, while cross-encoders have to learn an entirely dataset-specific classifier. The resulting decision rule in generative matchers preserves broader priors about entity identity, making it more reusable across datasets.

\subsection{Transfer Depends on Both Target and Source Data}
\label{sec:cross:effects}

Figure~\ref{fig:cross-dataset-retention}(b) reports the complete result matrices, where rows denote training source dataset and columns denote evaluation target datasets. The \texttt{OUT} column gives the mean off-diagonal retention of each source dataset, while the \texttt{IN} row gives the mean retention received by each target dataset. The transfer matrices show that transferability depends on both the evaluation target and the fine-tuning source. On the receiving side, some targets benefit substantially from external supervision. This pattern is most pronounced for \texttt{Semi-Heter}, whose mean incoming retention reaches 141.2 for cross-encoders and 104.3 for generative matchers. On average, models trained on other datasets therefore outperform models fine-tuned specifically on \texttt{Semi-Heter}. A similar but weaker effect appears for \texttt{Semi-Rel}, where incoming retention reaches 101.4 for generative matchers and 95.6 for cross-encoders. These results indicate that target-specific fine-tuning does not always provide enough supervision, particularly when the target evaluation distribution is weakly represented in its own training split.

Transferability also varies across training sources. For both cross-encoders and generative matchers, \texttt{WDC20} and \texttt{WDC80} emerge as particularly strong sources. These datasets are derived from the WDC product-matching benchmark and contain many challenging corner cases. Their strong outgoing retention suggests that exposure to such difficult and diverse matching patterns yields supervision that transfers effectively across datasets.
Source- and target-side transfer therefore reveal complementary aspects of what fine-tuning learns: the former measures how reusable the learned behavior is, while the latter shows where externally learned behavior remains effective.

\section{Cost Analysis}
\label{sec:cost}

Matcher architectures differ not only in predictive quality but also in their computational requirements, making effectiveness alone insufficient for practical model selection. We therefore complement the performance analysis with a systematic comparison of training and inference costs. 
We derive training costs from the logs recorded during the main experimental runs. For inference-cost evaluation, we restrict the generative matcher to producing a single output token for a fair comparison. This choice is motivated by our observation that the fine-tuned model consistently produces a valid response as its first generated token, making additional decoding unnecessary.

\headerul{Training cost} We compare runtime and peak GPU memory across architectures, averaged over datasets and seeds. Bi-encoders require the longest training time at every model size: their mean runtime is 1.9, 1.7, and 1.5 times that of the corresponding cross-encoder at 0.6B, 4B, and 8B, respectively. This overhead arises because each training pair requires two independent record encodings. Cross-encoder and generative runtimes are much closer, differing by less than 4.0\% at 4B and 8B.
Generative training is instead distinguished by memory consumption. At 8B, mean peak memory is 67.4\,GiB, compared with 48.7\,GiB for the cross-encoder and 41.1\,GiB for the bi-encoder; the largest generative run reaches 88.1\,GiB. Thus, relative to cross-encoders, the main additional cost of generative training is memory rather than runtime, largely due to the language-modeling head.

\begin{figure}[t]
\centering
\includegraphics{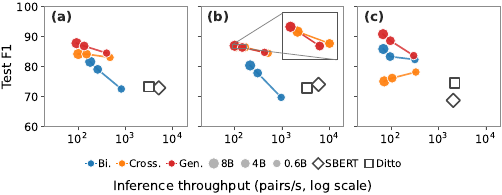}
\caption{Effectiveness--efficiency frontiers. The panels report results on all nine datasets, the seven datasets excluding \texttt{sh} and \texttt{sr}, and \texttt{sh} and \texttt{sr}, respectively. Throughput is averaged over five runs using the largest device-feasible batch size.}
\label{fig:cost-quality}
\end{figure}

\headerul{Inference effectiveness and throughput} For each \seqsplit{architecture-model} size configuration, we measure throughput using the largest feasible batch size found by starting from a large value and repeatedly halving it after an out-of-memory failure. Each configuration is run five times, and we report the mean throughput. Figure~\ref{fig:cost-quality} plots test F1 against candidate pairs processed per second.
Across all datasets, larger models generally trade throughput for effectiveness. The 0.6B cross-encoder already achieves an F1 score of 83.0 at approximately 473 pairs/s, whereas the 8B bi-encoder reaches 81.5 at only 181 pairs/s. On the seven datasets excluding \texttt{Semi-Heter} and \texttt{Semi-Rel}, the 4B cross-encoder occupies a favorable region of the frontier, achieving 86.4 F1 at approximately 165 pairs/s, while scaling to 8B yields little additional accuracy. On \texttt{Semi-Heter} and \texttt{Semi-Rel}, generative matchers provide the strongest effectiveness that is followed by bi-encoders.

\headerul{When bi-encoder caching pays off} The pairwise throughput comparison understates the principal systems advantage of bi-encoders: record representations can be computed once and reused during blocking. We consider a setting in which each record from Source A retrieves \(k\) candidate records from Source B. The Source A representation is then reused across its \(k\) candidate pairs.
Table~\ref{tab:results_by_k} estimates the time required for one million candidate decisions under different blocking choices. As \(k\) increases, the bi-encoder encoding cost is amortized across more candidates, whereas cross-encoder and generative costs remain pair-dependent. At 8B, the estimated bi-encoder time decreases from 16,931.9 seconds at \(k=1\) to 1,693.2 seconds at \(k=10\) and 169.3 seconds at \(k=100\). Bi-encoders are therefore most attractive when blocking retrieves multiple candidates per source record, rather than used as pairwise classifiers without caching.

\begin{table}[t]
\centering
\caption{Estimated inference time for 1M candidate-pair decisions in seconds. For the bi-encoder, \(k\) is the number of Source B candidates retrieved for each Source A record during blocking.}
\label{tab:results_by_k}
\resizebox{\columnwidth}{!}{%
\begin{tabular}{c cccc cc}
    \toprule
    \multirow{2}{*}{Model}
    & \multicolumn{4}{c}{Bi-encoder}
    & \multirow{2}{*}{Cross-encoder}
    & \multirow{2}{*}{Generative} \\
    \cmidrule(lr){2-5}
    & \(k=1\) & \(k=5\) & \(k=10\) & \(k=100\) & & \\
    \midrule
    0.6B & \phantom{0}3129.4 & \phantom{0}625.9 & \phantom{0}312.9 & \phantom{0}31.3 & \phantom{0}3130.7 & \phantom{0}3134.4 \\
    4B   & 11118.5 & 2223.7 & 1111.9 & 111.2 & 11113.6 & 11107.4 \\
    8B   & 16931.9 & 3386.4 & 1693.2 & 169.3 & 16920.5 & 16949.2 \\
    \bottomrule
\end{tabular}%
}
\end{table}

\begin{figure}[t]
\centering
\includegraphics[width=0.9 \columnwidth]{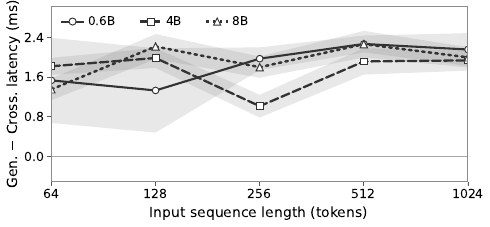}
\caption{Single-request latency overhead of the generative matcher relative to the cross-encoder as input length increases.}
\label{fig:latency-gap}
\end{figure}

\headerul{Latency comparison between cross-encoders and generative matchers} We further compare the per-query inference latency of cross-encoder and generative matchers using entity-pair inputs of equal length. For each model size and input length, we measure single-query latency over 100 independent runs. Because the two architectures use the same joint input and decoder backbone, their latency difference is primarily attributable to the output head and decoding procedure. Figure~\ref{fig:latency-gap} shows that the cross-encoder is consistently faster across all model sizes. The absolute gap remains approximately 2\,ms per query and becomes less consequential as input length increases, because backbone computation accounts for an increasingly large fraction of total latency.

\section{Recommendations}
\label{sec:recommendations}

We distill our findings into concrete suggestions for EM practitioners, researchers, and benchmark designers.

\header{Guidelines for EM practitioners} The choice between commercial API inference and fine-tuning should depend on workload volume. API-based inference is attractive for low-volume or rapidly changing workloads because it avoids training and deployment overhead. For sustained workloads, the one-time fine-tuning cost can be amortized after a few thousand API calls under the pricing and hardware assumptions used in our analysis.
 For supervised pairwise EM with labeled data reasonably representing deployment conditions, a medium-scale cross-encoder is a strong default. In our experiments, a 4B model achieves comparable performance as the 8B model while providing higher throughput and lower memory cost (\S~\ref{sec:three-layer:interaction}, \S\ref{sec:cost}). 
 Generative matchers are preferable when robustness to distribution shift is a primary concern. Their advantage appears in deployment scenarios involving distribution shifts, such as unseen schema combinations or cross-dataset transfer (\S\ref{sec:cross}).
 Model scaling is less important than building representative and high-coverage fine-tuning datasets. Larger models do not consistently improve EM performance and may amplify reliance on misleading matching cues (\S~\ref{sec:three-layer:generalization}).
Bi-encoders are suitable for settings in which record representations can be cached and reused after retrieval and blocking, providing competitive performance with higher throughput (\S~\ref{sec:cost}).

\header{Suggestions for EM researchers} Rather than pursuing incremental gains on saturated within-dataset benchmarks, future work should place greater emphasis on settings that directly evaluate model generalization, such as cross-dataset transfer (\S\ref{sec:cross:trans}). Researchers should investigate which matching signals models actually learn and whether these signals remain reliable across distributions. Such analyses can reveal when models rely on transferable evidence and when they exploit dataset-specific shortcuts~(\S~\ref{sec:cross:effects}).

\header{Design principles for EM benchmarks} Benchmark splits should explicitly characterize changes in attribute combinations, source distributions, and attribute-label relationships across training and evaluation data. Benchmark evaluation should move beyond reporting aggregate F1 and provide finer-grained analyses of model behavior, including precision, recall, and performance on relevant subgroups and difficult cases.
\section{Related Work}
\label{sec:related}

We review prior work along two dimensions: advances in neural and language model-based entity matching, and benchmarks and experimental studies that examine how evaluation design shapes empirical conclusions.

\header{Neural and language model-based entity matching} Conventional supervised entity matching relies on manually designed similarities and classifiers~\cite{konda2018magellan, whang2014incremental, tang2025description}. Neural methods replace these features with learned representations and comparison functions. DeepMatcher~\cite{mudgal2018deep} organizes this design space around attribute representation, comparison, and evidence aggregation, while DeepER and subsequent systems extend neural representations to blocking and matching~\cite{joty2018distributed,barlaug2021neural,papadakis2020blocking, krivosheev2023graph, tu2022domain}.

Pre-trained language models shifted the dominant formulation toward textual serialization and joint pair classification. Ditto fine-tunes a Transformer over the concatenated record pair and incorporates domain knowledge and augmentation~\cite{li2020deep,li2023effective}. Later work explores hierarchical attention, contrastive and self-supervised objectives, prompt tuning, augmentation, and interpretation~\cite{yao2022entity,peeters2022supervised,miao2021rotom,wang2023sudowoodo,wang2022promptem,paganelli2022analyzing}. A complementary line independently encodes records for reusable retrieval and blocking~\cite{thirumuruganathan2021deep, brinkmann2024sc}. Experimental studies show that the selected pretrained representation and fine-tuning strategy can substantially affect ER performance~\cite{zeakis2025depth,karapiperis2026impact}.

Recent work evaluates generative LLMs through zero-shot, few-shot prompting, generated rules task-specific fine-tuning, etc~\cite{peeters2025entity,steiner2025fine,wang2025match, hussein2025llm, zeakis2025avenger, yin2025talk, teofili2026can}. Multi-dataset training and LLM-generated supervision have also been used to build transferable smaller matchers~\cite{zhang2025deep,steiner2026labeling, zeakis2026distiller}. Given the high computational cost of language model-based matchers, a complementary line of work studies cost-aware system design~\cite{pulsone2026beacon, pulsone2026understanding, fan2024cost, balzotti2026entity}, toward scalable and automatable EM pipelines~\cite{maciejewski2025progressive, nikoletos2025auto, karapiperis2025sper}.

Another broad line of work reduces dependence on target-specific labels through active learning, weak supervision, pseudo-labeling, transfer learning, and domain adaptation~\cite{kasai2019low,kirielle2022transer,jin2021deep,trabelsi2022dame, wu2023ground, gagliardelli2024gsm}. Generalized and multi-dataset matchers similarly seek supervision that transfers across domains and schemas. Recent LLM studies evaluate transfer to unseen entities, datasets, and topical domains, showing that prompting and fine-tuning can behave differently under distribution shift~\cite{steiner2025fine,zhang2025deep, pulsone2026beacon}.

\header{Entity matching benchmarks and experimental analyses} ER benchmarks have enabled comparisons across domains and matcher families, but their conclusions depend on schema structure, candidate generation, label composition, entity overlap, and evaluation protocol. MaChAmp broadens evaluation beyond aligned relational tables to structured, semi-structured, and unstructured sources~\cite{cikm21machamp}. WDC Products varies training-set size, corner-case prevalence, and generalization to unseen entities~\cite{peeters2023wdc}. Other benchmark efforts examine schema and representation heterogeneity, open-world entities, imbalance, and more realistic integration settings~\cite{wangbridging,moslemi2026heterogeneity}.

Several E\&A studies analyze these evaluation conditions directly. Critical benchmark re-evaluations find that simple lexical or linear models remain competitive on many established datasets, limiting their ability to distinguish advanced systems~\cite{papadakis2024critical}. Experimental analyses of pretrained embeddings compare representations across blocking, supervised matching, and unsupervised matching~\cite{zeakis2025depth, brinkmann2024sc}, while fine-tuning studies examine when adaptation produces consistent gains~\cite{karapiperis2026impact}. Fairness analyses reveal subgroup differences hidden by aggregate scores~\cite{shahbazi2023through},  providing evidence for the development of fairer EM systems~\cite{araujo2026x, araujo2025treats}. 
Robustness studies expose sensitivity to perturbations and training-data composition~\cite{akbarian2022probing}. 
SMBench jointly benchmarks filtering and verification methods across real-world datasets, highlighting their interdependence in end-to-end entity-resolution performance~\cite{astappiev2026smbench}.
Karapiperis et al.~\cite{karapiperis2026impact} present the work most closely related to ours by examining the impact of fine-tuning on language model-based entity matching. However, their study exclude generative matchers and offers limited analysis of why matcher architectures differ in performance. Our work extends this line of research through a controlled comparison of matcher architecture, checkpoint objective, model scale, transferability, and computational cost within a single model family.
More broadly, these studies show that benchmark composition should be treated as an experimental factor rather than as neutral infrastructure.

\section{Conclusion}
\label{sec:discussion}

This work presents a controlled study of language model-based entity matching that disentangles matcher architecture from initialization with a model variant and model size. By comparing bi-encoders, cross-encoders, and generative matchers within a unified model family, we clarify which performance differences arise from representation quality, pairwise interaction, prediction formulation, and distribution shift rather than from architecture alone. Our results challenge the assumption that increasingly generative or larger models are universally superior and instead show that each matcher architecture is best suited to different data and deployment conditions. Beyond within-dataset accuracy, our study emphasizes transferability, robustness, and computational cost as essential complementary dimensions of evaluation. Overall, our findings provide a reliable basis for selecting matcher architectures, interpreting benchmark results, and designing future entity matching systems and evaluations.

\phem{Limitations} This study has several limitations that motivate future work. First, we keep the prompt, record serialization, and output label words (i.e., \texttt{yes} and \texttt{no}) fixed and therefore do not isolate their individual effects. Second, although we attribute the divergence between cross-encoder and generative matchers primarily to their prediction formulations, we do not directly isolate the contribution of the language-model head. Third, our analysis of the bi-encoder scaling is limited by the availability of larger embedding-oriented checkpoints, leaving open whether further scaling can close the gap with cross-encoders. Fourth, our throughput measurements use a heuristic search for the largest feasible batch size and may not identify the optimal serving configuration; specialized frameworks such as vLLM could provide stronger throughput but would introduce additional systems-level confounders. Finally, we focus on pairwise matching and leave the use of LLM-based bi-encoders for blocking and candidate generation to future work.

\bibliographystyle{ACM-Reference-Format}
\bibliography{sample}

\end{document}